\documentclass[11pt, letterpaper, twocolumn]{article}

\usepackage[margin=0.75in]{geometry}
\usepackage[utf8]{inputenc}         
\usepackage[T1]{fontenc}            
\usepackage{lmodern}                
\usepackage{microtype}              
\usepackage{amsmath, amssymb}       
\usepackage{graphicx}               
\usepackage{xcolor}                 
\usepackage{authblk}                
\usepackage[numbers, sort&compress]{natbib} 
\usepackage{booktabs}               
\usepackage{subcaption}
\usepackage{array}
\usepackage{booktabs}
\usepackage{siunitx}

\newcommand{\figref}[1]{Figure~\ref{#1}}
\newcommand{\tabref}[1]{Table~\ref{#1}}
\newcommand{\equref}[1]{Equation~(\ref{#1})}

\usepackage[
    colorlinks=true,
    linkcolor=blue!70!black,
    citecolor=blue!70!black,
    urlcolor=blue!70!black,
    breaklinks=true
]{hyperref}
\usepackage{xurl}
\usepackage{doi}

\title{\vspace{-2em} \textbf{Category differences in similarity between fixation density maps and psychological potential fields for sketches}}

\date{} 

\author[1]{Naoyuki Awano\thanks{Corresponding author: \texttt{awanonaoyuki@gmail.com}}}
\author[2]{Yuki Hayashi}

\affil[1]{\small Faculty of Business, Kokushikan University}
\affil[2]{\small Graduate School of Informatics, Osaka Metropolitan University}

\begin{document}

\maketitle

\begin{abstract}
When an object comes into view, humans can instantaneously recognize it if known to them. However, assessing how this recognition takes place remains challenging. A few studies using simple sketches have reported varying degrees of image similarity between the fixation density map (FDM), which visualizes eye movements and durations, and the psychological potential field (PPF), which visualizes part of the visual cognition process, depending on the object category. This study reports how the similarity behaves across three categories---animal, non-animal, and meaningless shape---and under rotated presentation, which is known to increase cognitive load during object recognition. The similarity was quantified through area under the curve (AUC) measures, including a baseline-adjusted AUC that isolates the early rise in similarity from stimulus-dependent baseline levels. A variability score was introduced to assess individual differences, and linear mixed models were used to evaluate category and orientation effects. Results indicated that the baseline-adjusted AUC yielded significant differences among all three categories with large effect sizes, whereas the raw AUC did not distinguish between the animal and non-animal categories due to the influence of sketch representation. No significant Category $\times$ Orientation interaction was observed, indicating that rotated presentation did not produce a category-specific effect. The meaningless shape stimuli yielded consistently low similarity and variability scores, clearly separating them from the two meaningful object categories. These results characterize the behavior of the FDM--PPF similarity under free viewing and indicate that baseline adjustment is necessary to separate category-related variation from stimulus-dependent baseline levels.
\end{abstract}

\section{Introduction}
Among studies on human sensory modalities, vision is considered our most important and complex sense \citep{hutmacher2019why}. Vision affects the processing of other senses and often takes precedence when integrating perceptual information \citep{winter2018vision}. Therefore, assessing and analyzing visual information is essential to understanding how people recognize objects.

Eye-tracking devices can measure visual behavior while placing a relatively low burden on participants. Eye tracking captures eye movements such as fixations and saccades, and has been reported to reveal pronounced trends related to cognitive processing \citep{kröger2020what}. Furthermore, attempts to measure eye movements using webcams have enabled large-scale online experiments \citep{jensen2022webcam,saxena2023deep}, making eye-tracking techniques promising for reducing both participant burden and experimental costs.

The most commonly measured eye movement is fixations, which record the location and duration of gaze \citep{tobii2024types}. Fixation points are recorded as coordinate sets, enabling quantitative cognitive analysis. Additionally, fixation points are visualized as fixation density maps (FDMs)---also known as fixation heatmaps or attention maps---which depict fixation locations and durations using a multicolor gradient, facilitating an intuitive understanding of gaze distribution. Overlaying an FDM onto the viewed image further supports qualitative analysis.

Eye fixations have also been studied in relation to object recognition and categorization. Gaze patterns during object viewing are known to reflect semantic content, with observers tending to fixate on behaviorally relevant regions such as faces in animals or functionally salient parts in artifacts \citep{kovic2009eyetracking,henderson2003human,hartendorp2013relation}. Studies using freehand sketches have suggested that fixation sequences may carry category-level structure even under a free-viewing condition without category priming \citep{sarvadevabhatla2017object}. However, accurate categorization from gaze patterns remains challenging.

On the other hand, there is the psychological potential field (PPF) \citep{yokose1970study}, a visualization theory of a visual perception phenomenon, and a classic theory in image processing. The PPF theory, defined through psychological experiments, describes a spatial field resembling an electrostatic field generated around a figure. Thus far, a computational model of the PPF has been established exclusively for binary images, and it is primarily regarded as a visualization or psychological image feature for simple images such as text, line drawings, and sketches. The PPF is an empirically defined field, and the perceptual mechanism it may reflect has not been established; in this study it is used as a stimulus-derived reference map rather than as a model of perception.

As both FDM and PPF visualize aspects of the visual perception process, an initial comparative analysis between them was conducted \citep{awano2020psyochological}. The results suggested that the similarity between FDM and PPF may vary depending on the type of stimulus object, with temporal changes in similarity scores showing trends akin to those observed in object categorization processes. However, these findings might be incidental, as the study relied on a limited set of ten stimulus images. Accordingly, whether these observations hold beyond that small stimulus set remains unclear.

This study reports the behavior of the FDM–PPF similarity on a larger and more structured set of stimuli. Using simple sketches as stimulus images---chosen both to simplify the problem and to satisfy the binary-image constraint of the PPF---we examined: (1) how the similarity scores differ among the primary categories of animal, non-animal, and meaningless shape; (2) how the scores change when stimuli are presented at rotated orientations, which are known to increase cognitive load during object recognition \citep{shepard1971mental,tarr1989mental,cutzu1994canonical,milivojevic2012object}; and (3) how the scores vary across individuals.

\section{Related works}
\subsection{Fixation density map (FDM)}
\label{subsec:fdm}
Eye tracking systems can measure various eye movements, including eye position, direction, pupil size, and gaze duration towards a point of interest \citep{klaib2021eye}. Numerous studies have used different eye tracking measures depending on their purpose, such as fixation positions, fixation duration, number of fixations, saccade amplitude, saccade duration, and blink rate \citep{zheng2022opportunities}. The most commonly measured elements are fixation positions and durations, visualized through an FDM, also known as a fixation heatmap or attention map. Areas viewed most frequently and for the longest durations are highlighted in prominent colors, enabling an intuitive understanding of fixation points and durations. Recently, it is noteworthy that machine learning has significantly enhanced the accuracy of FDM estimations, particularly in predicting visual saliency \citep{linardos2021deepgaze,lou2022transalnet,kümmerer2025modeling}.

There are four primary FDM visualization methods in the form of heatmaps: fixation count, absolute gaze duration, relative gaze duration, and participant percentage heatmaps (see \citep{bojko2009human} for further details). We utilized the absolute gaze duration heatmap as the FDM because it effectively represents the focused areas and duration of gaze. This choice was particularly appropriate for our experiment, where the viewing duration was standardized across all participants (see the \nameref{subsec:expmethd} section). The method for creating an FDM \citep{courtemanche2018physiological} involves preparing a blank image of the same size and resolution as the stimulus image or screen. Fixation points are then projected onto this blank image using a scaling function, typically a Gaussian function, centered on each gaze point, and subsequently accumulated. The scaling function parameters are determined based on viewing distance and visual angle. The most fixated locations are highlighted in a prominent color, while others are displayed in a gradient of less conspicuous colors.

\subsection{Psychological potential field (PPF)}
\label{subsec:ppf}
Drawing on studies of figural aftereffects \citep{köhler1944figural} and contour effects \citep{werner1935studies}, Yokose identified a spatial field similar to an electrostatic field surrounding figures \citep{yokose1970study}, which he termed the Psychological Potential Field (PPF), also known as the induction field in vision or field of visual perception. The PPF posits that when observing a figure, we perceive both the figure itself and the surrounding ripple effects. The PPF theory indicates that while the background and object colors influence it, it is primarily affected by object contours \citep{akagawa2003effect,ando2007effect}. However, despite the proven existence of PPF \citep{kaji1974contour,fukouzu1998analysis}, numerous aspects of its influence remain elusive. Consequently, the computational model of PPF applicable to digital images has been developed exclusively for binary black-and-white images.

In the PPF theory applied to digital images, the PPF spatial field is defined by calculating potential values for all background pixels. The potential value $M_p$ for a background pixel $\boldsymbol{p}$ is calculated as follows:
\begin{equation}
\label{eq1}
    M_p = \frac{1}{|C(\boldsymbol{p})|}\sum_{\boldsymbol{c}\in C(\boldsymbol{p})}{\frac{1}{d(\boldsymbol{p},\boldsymbol{c})}},
\end{equation}
where $C(\boldsymbol{p})$ represents the set of foreground contour pixels illuminated and not obscured by other foreground pixels when a light source is positioned at $p$, and $d(,)$ is the distance function.

There are two types of visualization methods for the PPF: heatmap and contour map. The PPF heatmap is visualized similarly to the FDM, employing color to denote different potential values. However, to facilitate intuitive understanding, potential values are accentuated using the following equation, given that pixel values near the figure change exponentially:
\begin{equation}
\label{eq2}
    M_p' = min(1.0, \alpha{M_p}/{M_{max}}),
\end{equation}
where $\alpha$ represents a bias factor, and $M_{max}$ is the maximum of all potential values. Conversely, the PPF contour map visualizes the field's distribution by depicting equipotential lines that connect pixels with identical potential values. Equipotential lines are frequently utilized not merely for visualization but also for the quantitative evaluation of design and impressions \citep{nagaishi2003evaluation}. In particular, analyzing the shape of these lines proves useful for arranging characters legibly \citep{miyoshi2002design}, designing arch bridges \citep{onaga1996relation}, and evaluating impressions of faces and hairstyles \citep{awano2018objective,awano2020complexity}. These applications demonstrate that the PPF has served as an effective evaluation tool for binary images of varying structural complexity, from simple characters to detailed facial contours.

\subsection{Similarity of FDM and PPF}
\label{subsec:simfdmppf}
FDM and PPF are both utilized as visualization techniques in visual perception. As they are distributive images represented as heatmaps, an initial comparative analysis between them was conducted \citep{awano2020psyochological}. Under conditions where the same resolution stimulus images were used and corresponding pixels were paired, the similarity between FDM and PPF was evaluated throughout the viewing duration. The similarity was quantified using the correlation coefficient, considered the fairest metric for comparison with FDMs \citep{bylinskii2019what}. However, given that PPF assigns no value to foreground pixels and that the number of foreground pixels is minimal in the sketches used in the experiment, these pixels were excluded from the similarity assessment. The comparison revealed that FDM and PPF are distinct in their visualization, each representing different processes in visual perception. Additionally, the similarity may have been indicative of the object categorization process in visual cognition, as it varied according to the type of stimulus objects. Furthermore, in the similarity assessments with FDM, PPF demonstrated higher categorization performance than other similar binary image features \citep{awano2022object}. This suggests that FDM and PPF potentially offer quantitative assessments of object categorization; however, the analysis was based on a limited set of only ten stimulus images.

\section{Methods}
\subsection{Experimental methodology}
\label{subsec:expmethd}
\begin{figure}[thb]
    \centering 
    \includegraphics[width=0.48\textwidth]{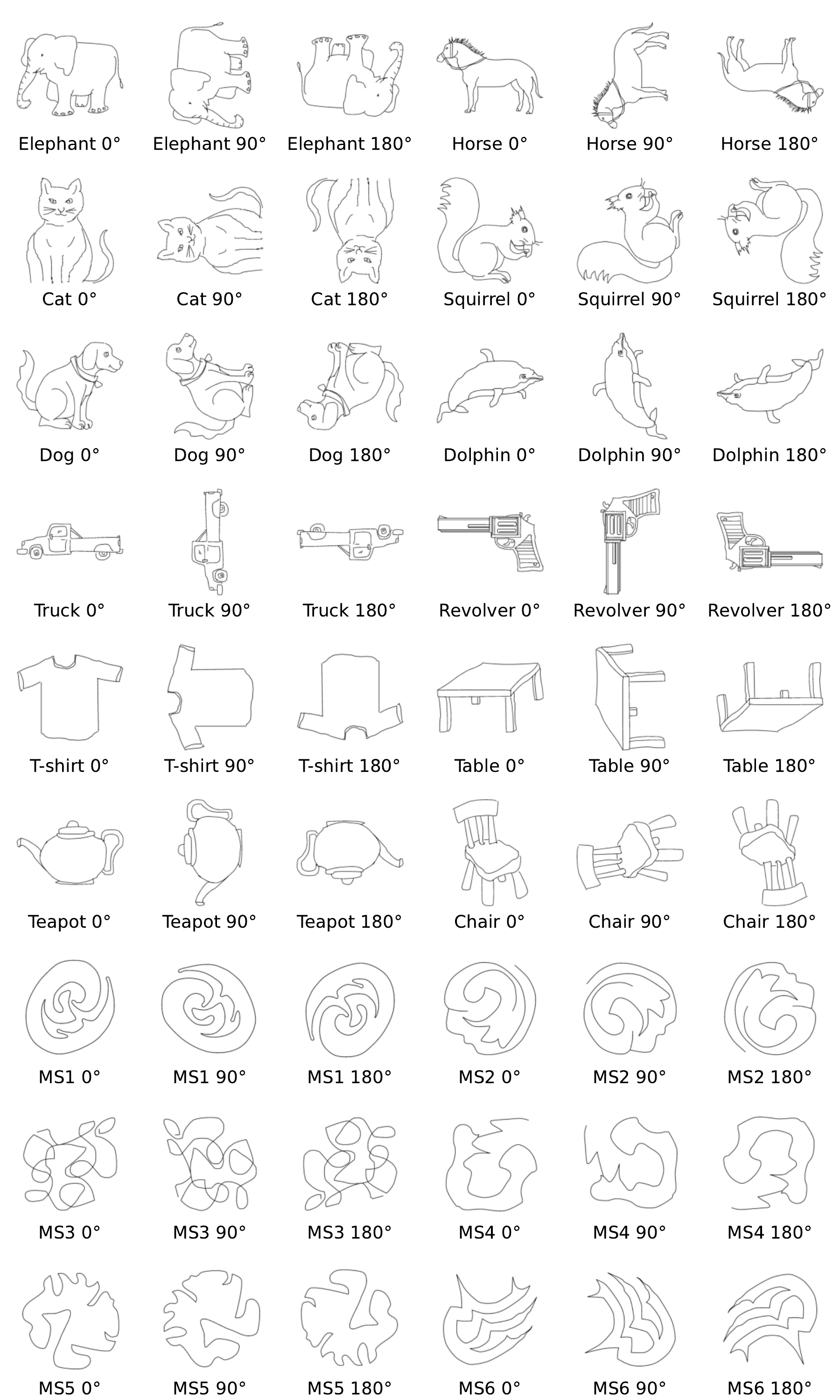}
    \caption{Stimulus images. The top three rows are the animal stimuli, the middle three are the non-animal stimuli, and the bottom three are the meaningless shape stimuli.} 
    \label{fig:stimuli}
\end{figure}

The study specifically examines the ability of similarity scores to categorize objects into the primary categories of animal, non-animal, and meaningless shapes. \figref{fig:stimuli} displays the stimulus images (see the first and fourth column images labeled as $0^{\circ}$). The animal and non-animal stimuli were sourced from a sketch dataset \citep{eitz2012hdhso}, and the meaningless shape stimuli were adopted from prior studies confirming that they are difficult to associate with nameable objects \citep{endo2003perceptual,aramaki2013naming}. Nevertheless, familiarity differences among individual sketches were not explicitly controlled.

Additionally, stimulus images were presented at rotated orientations ($90^{\circ}$ and $180^{\circ}$ counterclockwise) to impose additional cognitive load during object recognition, as shown in \figref{fig:stimuli}. If the similarity scores reflect both the object categorization and the recognition delay, this could suggest that the scores are indicative of cognitive processing. We chose stimulus images with a standard viewing orientation, except for the meaningless shape stimuli, which were then rotated $90^{\circ}$ and $180^{\circ}$ counterclockwise, as shown in \figref{fig:stimuli} (see images labeled $90^{\circ}$ and $180^{\circ}$).

\begin{figure}[tb]
    \centering 
    \includegraphics[width=0.48\textwidth]{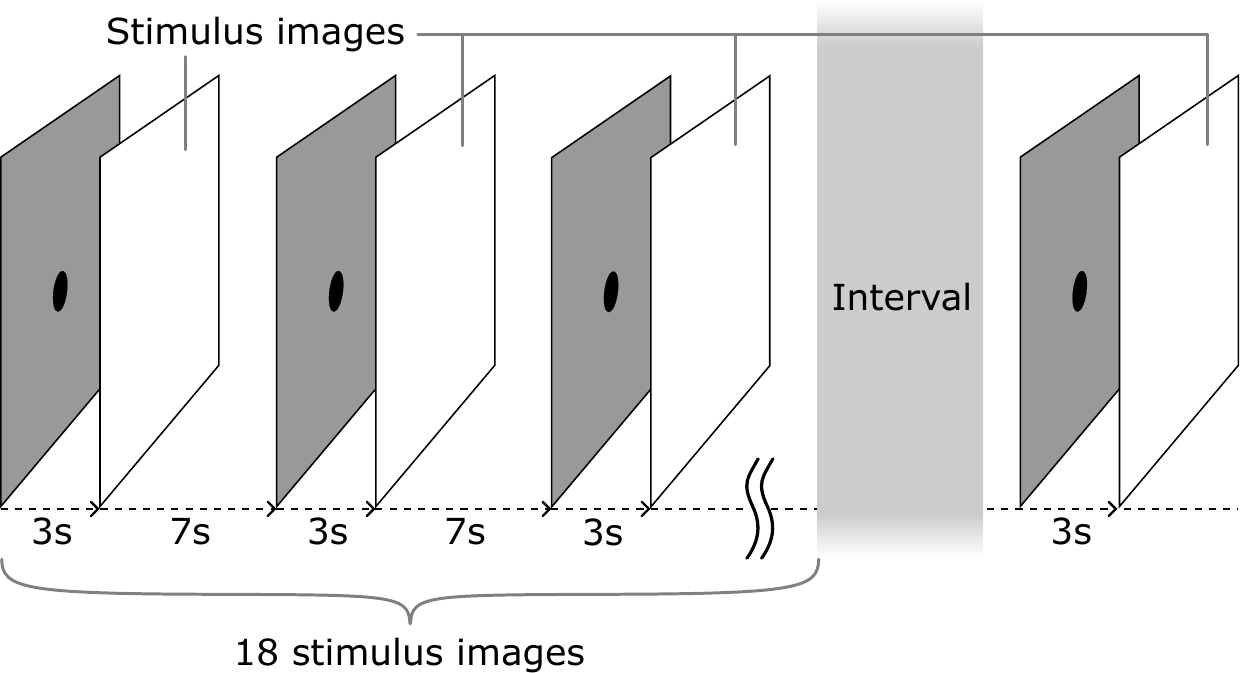}
    \caption{Stimulus image displaying method for the eye tracking experiment.} 
    \label{fig:procedure}
\end{figure}

An eye-tracking experiment was conducted with the 54 stimulus images in \figref{fig:stimuli} to generate the corresponding FDMs. Participants sat in front of a 24.1 inch display equipped with an eye tracker, stabilized their heads using a chin rest, and underwent calibration. According to \citep{courtemanche2018physiological}, the distance from the center of the display to the participants' eyes was set at 70 cm, with a foveal vision angle of $2^{\circ}$. Participants viewed each stimulus image freely for 7 seconds. A gray inter-stimulus screen featuring a central black circle was displayed for 3 seconds between each stimulus to reduce the afterimage effect from the previous stimulus. To account for participant fatigue, breaks were interspersed after every 18 stimulus images, with the duration of each interval being at the discretion of the participant, as illustrated in \figref{fig:procedure}. The sequence of the stimulus images was randomized, ensuring that the same image, albeit rotated, was not shown in succession. Twenty-five university students (mean age 20.2 years, SD = 1.3) participated in the study. Data from seven participants were excluded based on the following criteria: (1) more than 30\% of fixation points fell outside the display area, or (2) valid fixations were absent for more than 30\% of the 0–3 s viewing interval. Participants meeting either criterion were excluded, leaving 18 for analysis. All participants provided written informed consent prior to the experiment.

The FDMs were generated from the acquired fixation points using the method outlined in the \nameref{subsec:fdm} section. Similarly, PPFs were produced according to the procedure detailed in the \nameref{subsec:ppf} section. However, as PPF represents a static image feature, the PPFs of stimulus images rotated $90^{\circ}$ and $180^{\circ}$ in the rotated conditions are identical to those of the $0^{\circ}$ images when rotated back to their original orientation.

\subsection{Similarity score metric}
\label{subsec:similarity_score_metric}
This study utilizes the same similarity scoring method for FDM and PPF as employed in the previous study \citep{awano2020psyochological}, outlined in the \nameref{subsec:simfdmppf} section. Specifically, this method uses the correlation coefficient to calculate image similarity, excluding foreground pixels. Since this method utilizes the correlation coefficient as an indicator of image similarity rather than under a general interpretation, the value range may hinder an intuitive understanding of the similarity. Therefore, we redefine the similarity score $S$ for enhanced clarity and normalize it as follows:
\begin{equation}
\label{eq3}
    S(\boldsymbol{I}_{fdm},\boldsymbol{I}_{ppf}) = \frac{\sigma_{fdm-ppf} + \sigma_{fdm}\sigma_{ppf}}{2\sigma_{fdm}\sigma_{ppf}},
\end{equation}
where $\boldsymbol{I}_{fdm}$ represents an FDM image, $\boldsymbol{I}_{ppf}$ a PPF image, $\sigma_{fdm}$ and $\sigma_{ppf}$ the standard deviations of $\boldsymbol{I}_{fdm}$ and $\boldsymbol{I}_{ppf}$ respectively, and $\sigma_{fdm-ppf}$ the covariance between $\boldsymbol{I}_{fdm}$ and $\boldsymbol{I}_{ppf}$. This formulation is equivalent to a linear rescaling of the Pearson correlation coefficient from $[-1,1]$ to $[0,1]$, preserving ordinal relationships, where higher scores signify greater image similarity. The similarity score was computed for each participant individually by comparing the participant's FDM with the PPF, and the resulting scores were then averaged across participants for each stimulus. Since the similarity score varies over the viewing duration, we summarize each score curve $S(t)$ by its area under the curve. The AUC over the full analysis window is defined as
\begin{equation}
\label{eq4}
  \mathrm{AUC} = \sum_{i=0}^{30} S(t_i), \qquad t_i = 0.1\,i~\text{s},
\end{equation}
summed over the similarity scores sampled every 100 ms. In addition, we define the baseline-adjusted AUC (Adj. AUC) as
\begin{equation}
\label{eq5}
  \text{Adj.\ AUC} = \sum_{i=0}^{10} \bigl( S(t_i) - S(t_0) \bigr),
\end{equation}
where $S(t_0)$ denotes the similarity score of the first sample ($0-100$ ms). Subtracting $S(t_0)$ removes a stimulus-dependent baseline: at viewing onset, the FDM is dominated by the Gaussian kernel centered near the initial fixation position, which is expected to overlap with the PPF to a degree that depends on the spatial structure of the stimulus contours. Summing over $0-1$ s then isolates the early rise in similarity, during which visual cognition for object categorization typically occurs (within a few hundred milliseconds \citep{thorpe2009speed,xu2019ultrarapid}). 

\subsection{Variability score metric}
While PPF is a static visualization image, FDM varies as it is based on fixation points acquired through eye tracking. Consequently, the similarity score is influenced by the participants' fixations. However, since FDM represents the accumulated fixation points of all participants, it fails to capture individual differences, only reflecting the ``average'' fixation tendency \citep{bojko2009human}. To evaluate variability in addition to the similarity score, we have defined the variability score $V$ using the following formulas:
\begin{equation}
\label{eq6}
    V(\boldsymbol{I}_{fdm},\boldsymbol{I}_{ppf})=\sqrt{\frac{1}{|H|}\sum\limits_{h\in H}\big(S(\boldsymbol{I}^h_{fdm},\boldsymbol{I}_{ppf})-\mu\big)^2},
\end{equation}
\begin{equation}
\label{eq7}
    \mu=\frac{1}{|H|}\sum\limits_{h\in H}{S(\boldsymbol{I}^h_{fdm},\boldsymbol{I}_{ppf})},
\end{equation}
where $H=\{h_1,h_2,...\}$ denotes the set of participants, with $\boldsymbol{I}^h_{fdm}$ representing the FDM of participant $h$. A high variability score signifies significant differences in the similarity score among participants, indicating substantial individual variation. As with the similarity score, we also report the AUC of the variability score over the observation period, denoted $V'$, as a cumulative summary of individual differences for each stimulus.

\subsection{Statistical analysis}
\label{subsec:stats_analysis}
To assess the effects of object category and orientation on the similarity scores, linear mixed models (LMMs) \citep{pinheiro2000mixed} were fitted using restricted maximum likelihood estimation. Category (animal, non-animal, meaningless shape) and Orientation ($0^{\circ}$, $90^{\circ}$, $180^{\circ}$) were included as fixed effects, with their interaction term. Participant and Stimulus were included as crossed random effects, with Participant as the grouping factor and Stimulus as a variance component, thereby accounting for both individual differences and stimulus-specific variability. Pairwise comparisons between categories were conducted using Bonferroni correction for three comparisons. Paired Cohen's $d$ was computed from participant-level means as a measure of effect size, and 95\% confidence intervals are reported for all estimated coefficients. The significance level was set at $\alpha=.05$ for all inferential tests.

Model assumptions were assessed by inspecting the conditional residuals and the estimated random effects (Supplementary Figure S1 and S2). For the AUC model, the residuals were approximately symmetric with no systematic variance trend, and the participant variance component was estimated near zero, indicating negligible between-participant differences in the raw AUC. For the Adj. AUC model, the residuals showed right skew and variance increasing with fitted values, reflecting the floor-level scores of the meaningless shape stimuli. Because linear mixed models are generally robust to such distributional violations for fixed-effect inference \citep{schielzeth2020}, and because all pairwise category comparisons were confirmed by nonparametric Wilcoxon signed-rank tests on participant-level means (Supplementary Table S1), the LMM results are reported in the main text.

A sensitivity power analysis using G*Power 3.1 \citep{faulStatisticalPowerAnalyses2009} indicated that, with 18 participants, an alpha level of $.05$, and a power of $.80$, the minimum detectable effect size was Cohen's $f=0.31$ (medium to large).

\subsection{Stimulus property analysis}
To assess the extent to which the similarity measures are influenced by low-level visual properties of the stimuli, six image features were computed from each of the 18 base ($0^{\circ}$) stimulus images: the foreground pixel ratio; the bounding-box area of the figure relative to the image area (content area); the foreground pixel density within the bounding box (line density); the number of contour pixels (perimeter); the number of eight-connected components; and a bilateral symmetry score, defined as the maximum Jaccard overlap between the dilated figure mask and its left–right or up–down mirrored counterpart. As these features are invariant to rotated presentation, Spearman rank correlations were computed between each feature and the stimulus-level AUC and Adj. AUC averaged across orientations. This analysis is descriptive and is intended to characterize potential confounds rather than to support inferential claims. In addition, to examine whether the category differences in the Adj. AUC could be attributed to these features, the LMM described in the \nameref{subsec:stats_analysis} section was refitted with the standardized foreground pixel ratio and line density added as stimulus-level covariates.

\section{Results}
\label{sec:rslt}

\begin{figure*}[ptbh]
    \centering 
    \includegraphics[width=0.9\textwidth]{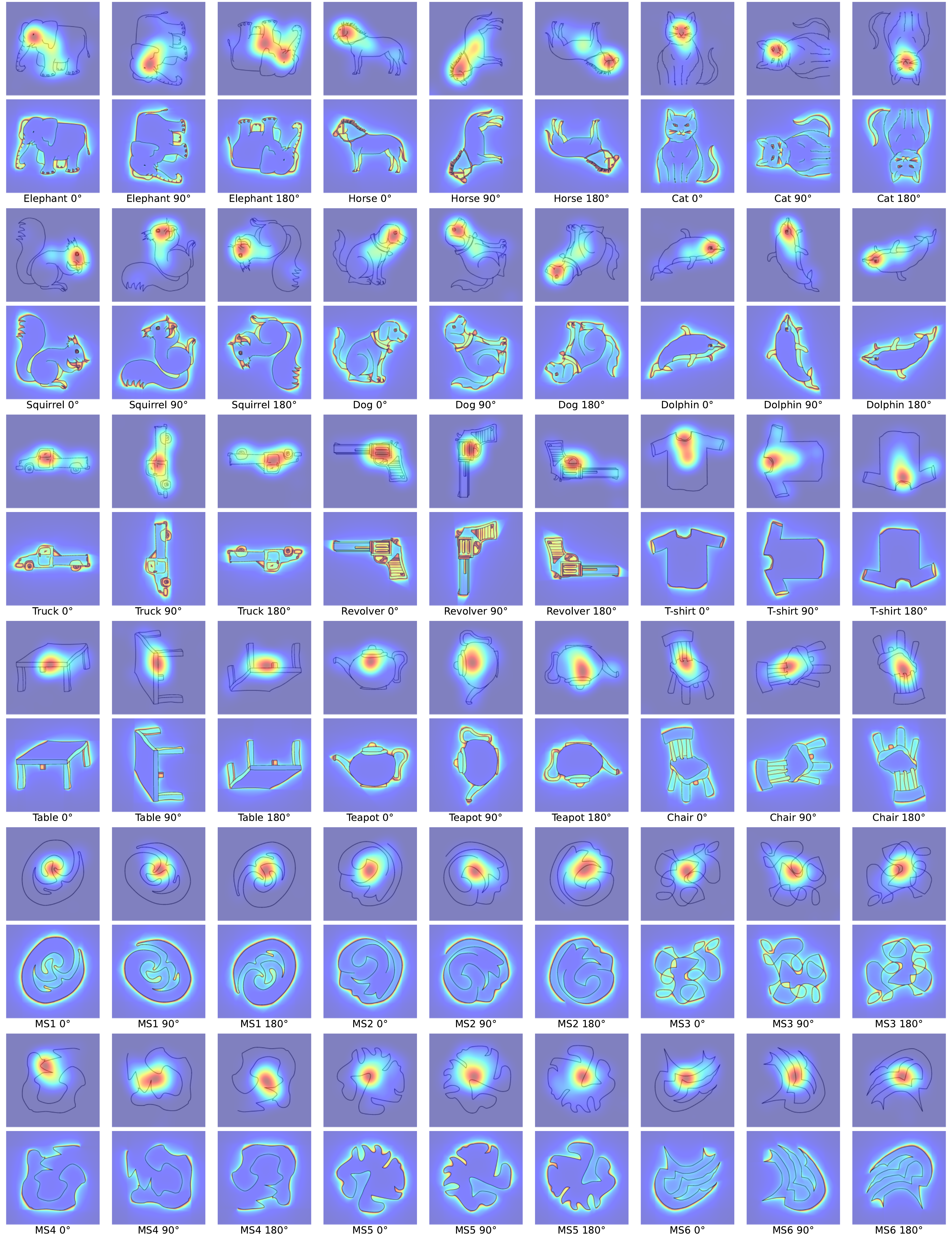}
    \caption{Visualization results (top: FDMs; bottom: PPFs). Each FDM is an absolute gaze duration heatmap accumulated over all participants, and each PPF depicts the potential values of \equref{eq1}, accentuated with the bias factor $\alpha=20$ in \equref{eq2} for visualization purposes only (all similarity and variability scores were computed from unbiased potential values, $\alpha=1$). In both maps, the color scale is normalized within each image, with red indicating the highest values—locations fixated most frequently and longest in FDMs, and high potential values in PPFs—and blue indicating the lowest.} 
    \label{fig:visualization}
\end{figure*}

\begin{figure*}[tbhp]
    \centering
    \begin{subfigure}[b]{0.32\linewidth}
        \centering
        \includegraphics[width=\linewidth]{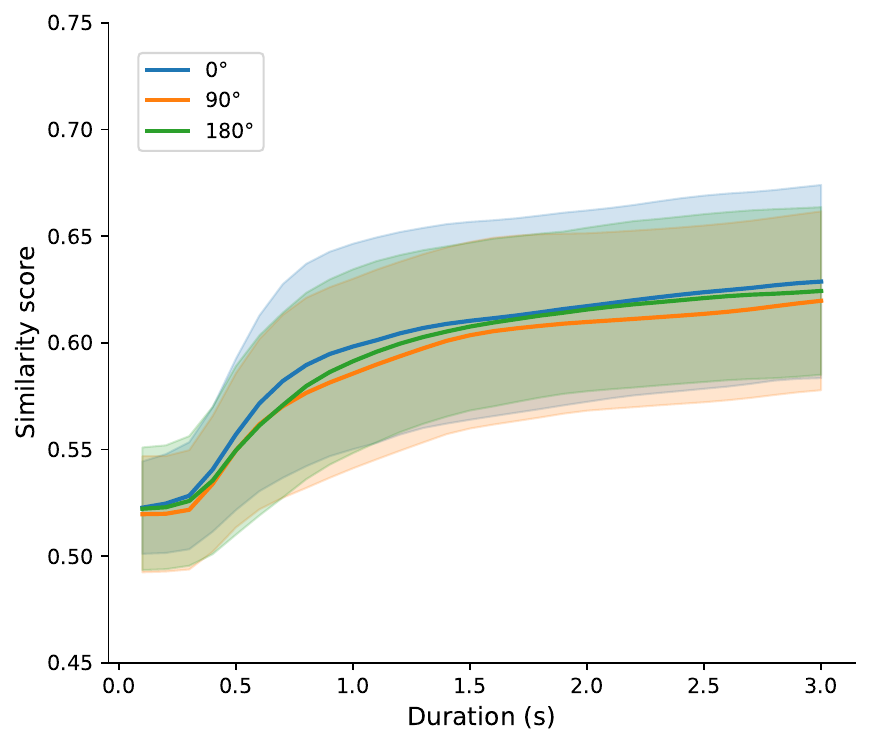}
        \caption{Animal}
    \end{subfigure}%
    \hfill
    \begin{subfigure}[b]{0.32\linewidth}
        \centering
        \includegraphics[width=\linewidth]{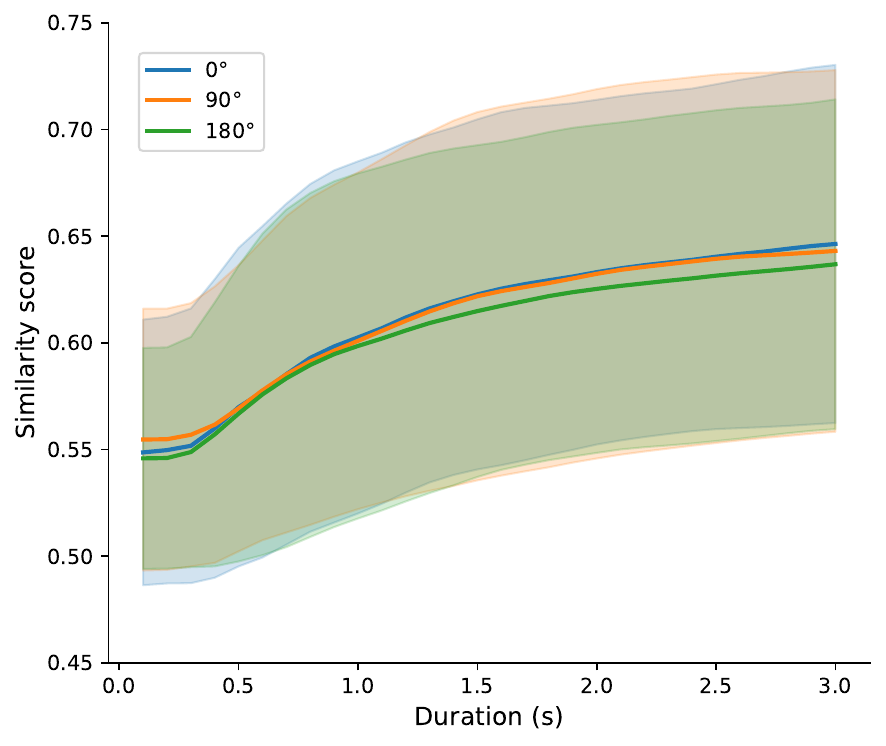}
        \caption{Non-animal}
    \end{subfigure}%
    \hfill
    \begin{subfigure}[b]{0.32\linewidth}
        \centering
        \includegraphics[width=\linewidth]{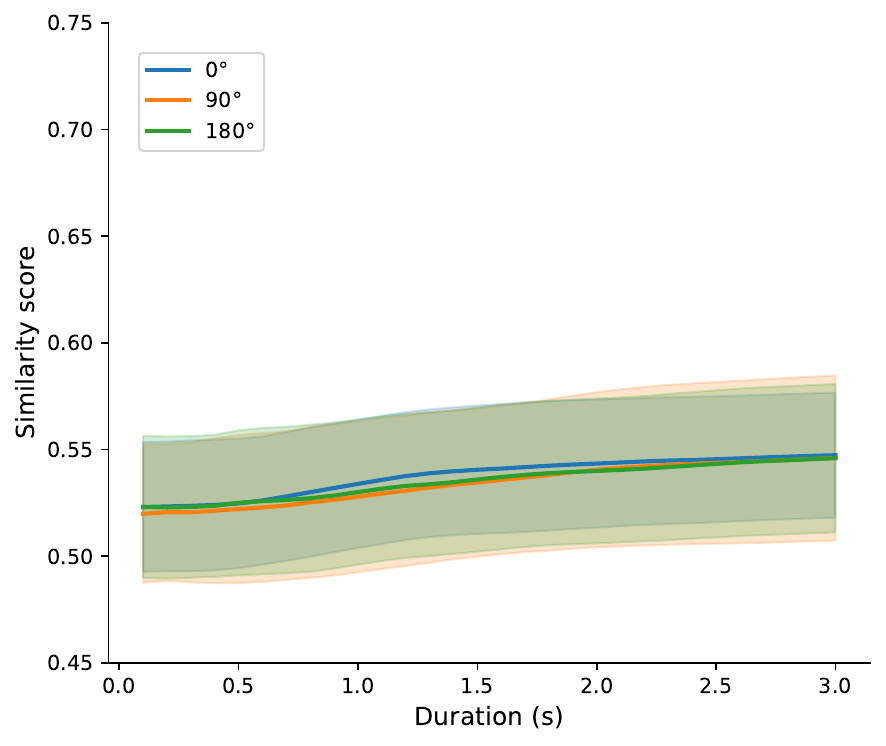}
        \caption{Meaningless shape}
    \end{subfigure}
    \caption{Transitions of the similarity scores in each object category. Each curve shows the similarity score $S(t)$ averaged over the six stimuli of the category at each orientation ($0^{\circ}$, $90^{\circ}$, and $180^{\circ}$), and the shaded bands indicate $\pm 1$ standard deviation across the six stimuli. Time is measured from stimulus onset.}
    \label{fig:similarity_category}
\end{figure*}

\begin{figure*}[tbhp]
    \centering 
    \includegraphics[width=0.85\textwidth]{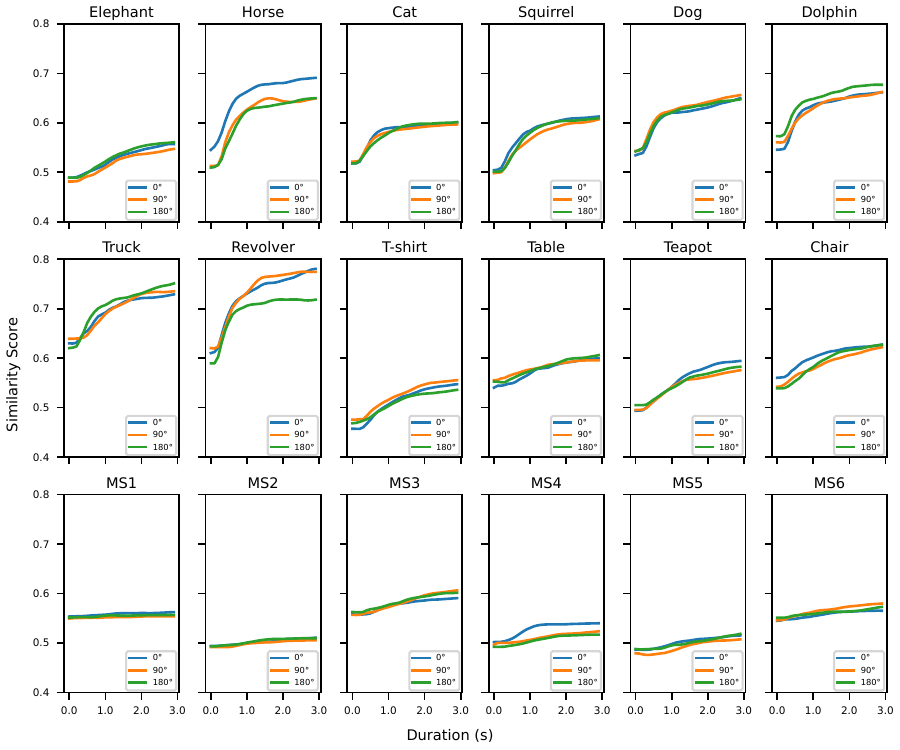}
    \caption{Transitions of the similarity scores in each stimulus.} 
    \label{fig:similarity_stimulus}
\end{figure*}

As outlined in the previous section, a duration of 7 seconds was deemed sufficient for generating FDMs. Nonetheless, the results presented here are based on fixation points collected within the initial 3 seconds. The reasons for this are threefold: visual cognition for object categorization typically occurs within a few hundred milliseconds \citep{thorpe2009speed,xu2019ultrarapid}; fixation points up to 3 seconds have commonly been utilized for FDM generation in other datasets \citep{bethge2023mit}; and the trends in our data did not show significant changes after the initial 3 seconds. \figref{fig:visualization} shows the visualization results for both FDMs and PPFs. Given that the PPF is a static image attribute, its potential values remain unchanged upon rotation; for example, a $90^{\circ}$ rotated PPF image aligns with the $0^{\circ}$ PPF image when it is rotated back $90^{\circ}$ clockwise. Note that, within \equref{eq2}, the bias factor was set to $\alpha=20$ exclusively for PPF visualization, following \citep{awano2020psyochological}, and $\alpha=1$ for the calculation of similarity and variability scores. This distinction ensures that the visualization in \figref{fig:visualization} clearly highlights the spatial distribution of potential values, while the similarity and variability scores are computed from unbiased potential values.

\begin{figure*}[t]
    \centering 
    \includegraphics[width=0.95\textwidth]{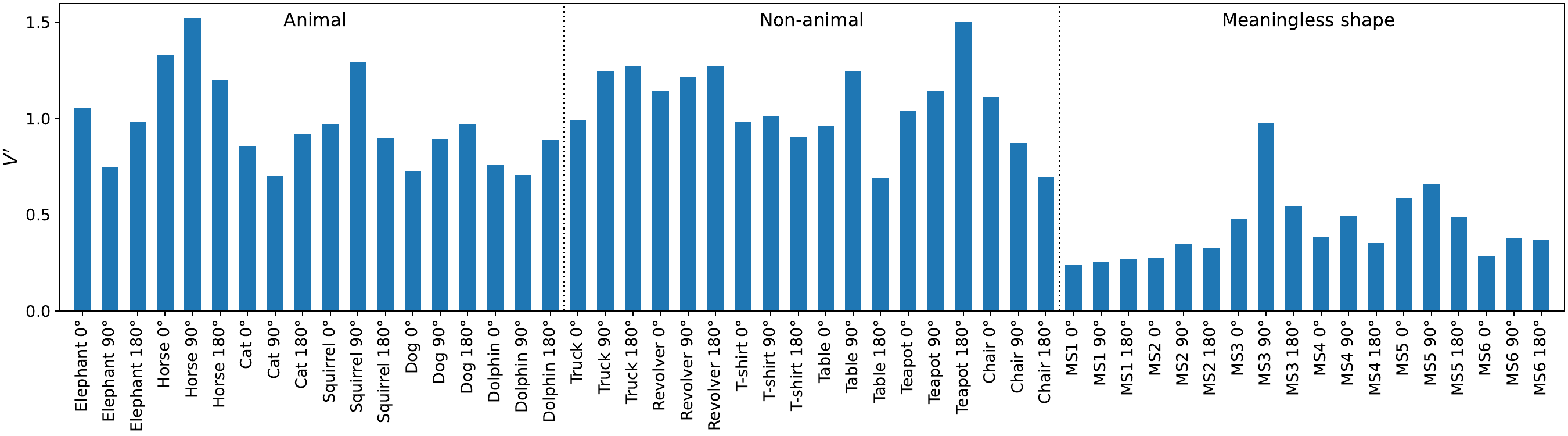}
    \caption{Total variability score $V'$ for each stimulus.} 
    \label{fig:total_variability}
\end{figure*}

\begin{figure*}[t]
    \centering 
    \includegraphics[width=0.95\textwidth]{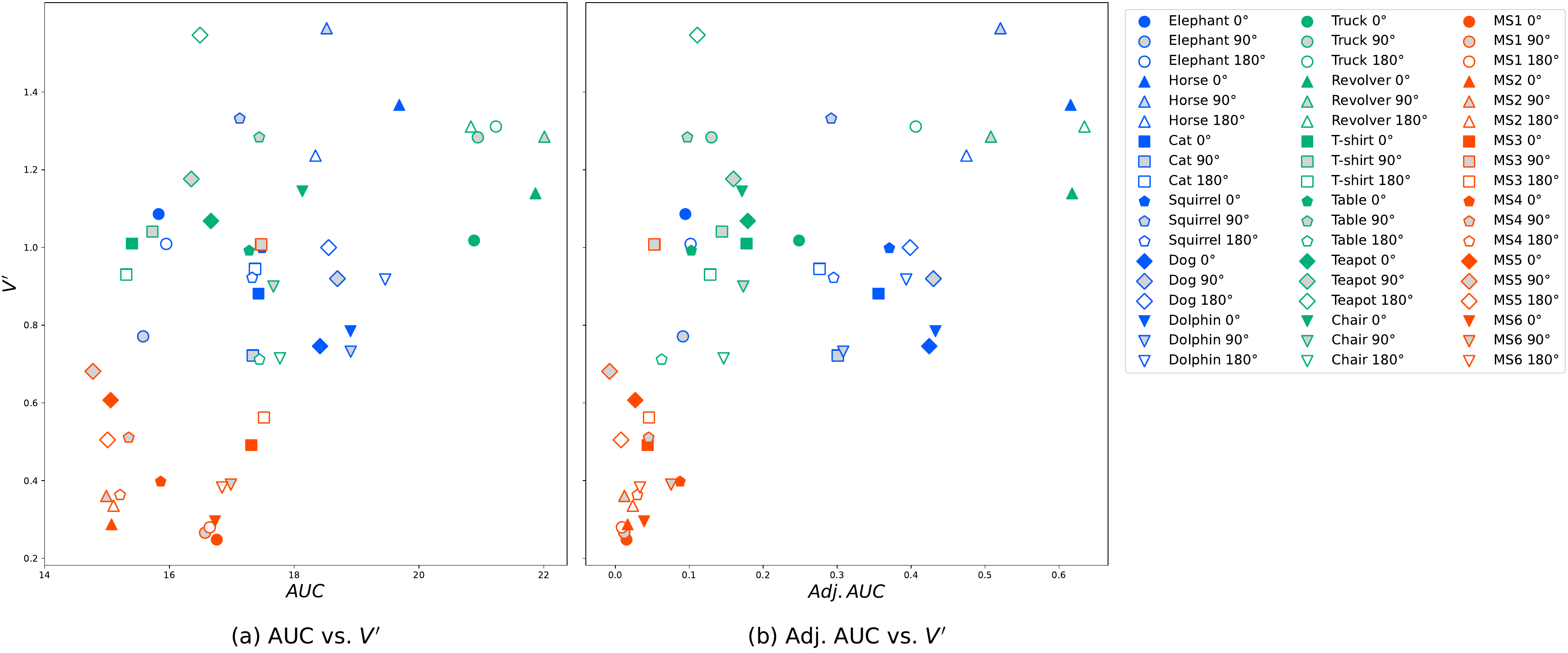}
    \caption{Relation between the total similarity score and total variability score $V'$ for each stimulus. (a) AUC of the similarity score over 0–3 s. (b) Baseline-adjusted AUC over 0–1 s. Blue markers represent animal stimuli, green for non-animal stimuli, and red for meaningless shape stimuli. In both panels, the meaningless shape stimuli cluster in the lower-left region, indicating consistently low similarity and variability.} 
    \label{fig:sv_relation}
\end{figure*}

\figref{fig:similarity_category} shows each category's similarity scores and corresponding standard deviations. For instance, the $0^{\circ}$ in \figref{fig:similarity_category}(a) displays the mean similarity scores and standard deviations for six animal stimuli: Elephant, Horse, Cat, Squirrel, Dog, and Dolphin. \figref{fig:similarity_stimulus} depicts the individual similarity scores for each stimulus. As these scores derive from correlation coefficients, associated p-values were also computed. \tabref{tab:results_compact}  presents the AUC and Adj. AUC for each stimulus. Additionally, \figref{fig:total_variability} shows $V'$ for each stimulus, and \figref{fig:sv_relation} presents the distribution of the AUC and Adj. AUC in relation to $V'$.

\tabref{tab:pairwise} summarizes the pairwise category comparisons from the LMMs. For the AUC, the animal and non-animal categories each differed significantly from the meaningless shape category (both $p<.001$, Bonferroni-corrected), whereas the difference between animal and non-animal did not reach significance after correction ($p=.108$). For the Adj. AUC, all three pairwise comparisons were significant (all $p<.001$, Bonferroni-corrected), with large effect sizes (all $|d|>1.0$). \tabref{tab:orientation} presents the orientation effects. Although some main effects of orientation reached significance, no significant Category $\times$ Orientation interactions were observed (all $p>.05$).

\begin{table}[htbp]
\centering
\fontsize{8.5pt}{8.5pt}\selectfont
\setlength{\tabcolsep}{3pt}
\caption{AUC of the similarity score over 0–3 s and Adj. AUC over 0–1 s for each stimulus. Background shading reflects relative magnitude within each column.}
\label{tab:results_compact}
\newlength{\mycolwd}
\settowidth{\mycolwd}{Adj. AUC}
\begin{tabular}{llc >{\centering\arraybackslash}p{\mycolwd} >{\centering\arraybackslash}p{\mycolwd}}
\toprule
Category & Stimulus & Orientation & AUC & Adj. AUC \\
\midrule
Animal & Elephant & 0$^\circ$ & 15.826 & 0.095 \\
  &   & 90$^\circ$ & 15.578 & 0.091 \\
  &   & 180$^\circ$ & 15.947 & 0.102 \\
\addlinespace
  & Horse & 0$^\circ$ & 19.683 & 0.616 \\
  &   & 90$^\circ$ & 18.519 & 0.521 \\
  &   & 180$^\circ$ & 18.340 & 0.475 \\
\addlinespace
  & Cat & 0$^\circ$ & 17.426 & 0.356 \\
  &   & 90$^\circ$ & 17.336 & 0.301 \\
  &   & 180$^\circ$ & 17.371 & 0.276 \\
\addlinespace
  & Squirrel & 0$^\circ$ & 17.483 & 0.371 \\
  &   & 90$^\circ$ & 17.125 & 0.292 \\
  &   & 180$^\circ$ & 17.325 & 0.295 \\
\addlinespace
  & Dog & 0$^\circ$ & 18.413 & 0.425 \\
  &   & 90$^\circ$ & 18.691 & 0.430 \\
  &   & 180$^\circ$ & 18.550 & 0.399 \\
\addlinespace
  & Dolphin & 0$^\circ$ & 18.902 & 0.433 \\
  &   & 90$^\circ$ & 18.905 & 0.308 \\
  &   & 180$^\circ$ & 19.457 & 0.393 \\
\midrule
Non-animal & Truck & 0$^\circ$ & 20.881 & 0.248 \\
  &   & 90$^\circ$ & 20.942 & 0.130 \\
  &   & 180$^\circ$ & 21.232 & 0.406 \\
\addlinespace
  & Revolver & 0$^\circ$ & 21.866 & 0.618 \\
  &   & 90$^\circ$ & 22.009 & 0.508 \\
  &   & 180$^\circ$ & 20.832 & 0.634 \\
\addlinespace
  & T-shirt & 0$^\circ$ & 15.400 & 0.177 \\
  &   & 90$^\circ$ & 15.727 & 0.144 \\
  &   & 180$^\circ$ & 15.307 & 0.128 \\
\addlinespace
  & Table & 0$^\circ$ & 17.277 & 0.103 \\
  &   & 90$^\circ$ & 17.438 & 0.097 \\
  &   & 180$^\circ$ & 17.442 & 0.063 \\
\addlinespace
  & Teapot & 0$^\circ$ & 16.664 & 0.179 \\
  &   & 90$^\circ$ & 16.349 & 0.160 \\
  &   & 180$^\circ$ & 16.489 & 0.111 \\
\addlinespace
  & Chair & 0$^\circ$ & 18.130 & 0.171 \\
  &   & 90$^\circ$ & 17.666 & 0.173 \\
  &   & 180$^\circ$ & 17.770 & 0.147 \\
\midrule
Meaningless & MS1 & 0$^\circ$ & 16.758 & 0.015 \\
shape &   & 90$^\circ$ & 16.571 & 0.012 \\
  &   & 180$^\circ$ & 16.645 & 0.009 \\
\addlinespace
  & MS2 & 0$^\circ$ & 15.071 & 0.017 \\
  &   & 90$^\circ$ & 14.988 & 0.012 \\
  &   & 180$^\circ$ & 15.103 & 0.024 \\
\addlinespace
  & MS3 & 0$^\circ$ & 17.313 & 0.044 \\
  &   & 90$^\circ$ & 17.470 & 0.053 \\
  &   & 180$^\circ$ & 17.515 & 0.045 \\
\addlinespace
  & MS4 & 0$^\circ$ & 15.859 & 0.088 \\
  &   & 90$^\circ$ & 15.346 & 0.045 \\
  &   & 180$^\circ$ & 15.207 & 0.030 \\
\addlinespace
  & MS5 & 0$^\circ$ & 15.057 & 0.027 \\
  &   & 90$^\circ$ & 14.773 & -0.008 \\
  &   & 180$^\circ$ & 15.008 & 0.008 \\
\addlinespace
  & MS6 & 0$^\circ$ & 16.730 & 0.039 \\
  &   & 90$^\circ$ & 16.984 & 0.075 \\
  &   & 180$^\circ$ & 16.843 & 0.033 \\
\bottomrule
\end{tabular}
\end{table}

\begin{table*}[htbp]
\caption{Pairwise category comparisons from linear mixed models with Bonferroni correction. AUC denotes the area under the similarity score curve over 0--3~s; Adj. AUC denotes the baseline-adjusted AUC over 0--1~s. Cohen's $d$ (paired) is computed from participant-level means.}
\label{tab:pairwise}
\centering
\fontsize{10pt}{10pt}\selectfont
\renewcommand{\arraystretch}{1.2} 
\setlength{\tabcolsep}{2.5pt} 

\begin{tabular}{
    l
    S[table-format=-1.2]
    c
    S[table-format=<.3]
    S[table-format=-1.2]
    S[table-format=-1.2]
    c
    S[table-format=<.3]
    S[table-format=-1.2]
}
\toprule
& \multicolumn{4}{c}{AUC} & \multicolumn{4}{c}{Adj. AUC} \\
\cmidrule(lr){2-5} \cmidrule(lr){6-9}
Comparison & {Coef.} & {95\% CI} & {$p$ (Bonf.)} & {$d$} & {Coef.} & {95\% CI} & {$p$ (Bonf.)} & {$d$} \\
\midrule
Animal vs. Meaningless shape    & -1.76 & [$-2.20, -1.31$] & <.001 &  4.41 & -0.31 & [$-0.36, -0.27$] & <.001 &  3.38 \\
Animal vs. Non-animal      &  0.47 & [$\phantom{-}0.03, \phantom{-}0.92$]   & .108  & -1.17 & -0.11 & [$-0.16, -0.06$] & <.001 &  1.53 \\
Non-animal vs. Meaningless shape & -2.23 & [$-2.68, -1.79$] & <.001 &  5.28 & -0.20 & [$-0.25, -0.16$] & <.001 &  3.00 \\
\bottomrule
\end{tabular}
\end{table*}

\begin{table*}[t]
\caption{Orientation effects from additive linear mixed models. Although some main effects reached significance, no significant Category $\times$ Orientation interactions were observed in the full models (all $p > .05$), indicating that the effect of rotated presentation did not differ systematically across categories.}
\label{tab:orientation}
\centering
\fontsize{10pt}{10pt}\selectfont
\renewcommand{\arraystretch}{1.2}
\setlength{\tabcolsep}{5pt}

\begin{tabular}{
    l
    S[table-format=-1.2]
    c
    S[table-format=-1.2]
    S[table-format=.3]
    S[table-format=-1.2]
    c
    S[table-format=-1.2]
    S[table-format=.3]
}
\toprule
& \multicolumn{4}{c}{AUC} & \multicolumn{4}{c}{Adj. AUC} \\
\cmidrule(lr){2-5} \cmidrule(lr){6-9}
Comparison & {Coef.} & {95\% CI} & {$z$} & {$p$} & {Coef.} & {95\% CI} & {$z$} & {$p$} \\
\midrule
$90^\circ$ vs.\ $0^\circ$  & -0.13 & [$-0.24, -0.02$] & -2.28 & .023 & -0.04 & [$-0.06, -0.01$] & -2.90 & .004 \\
$180^\circ$ vs.\ $0^\circ$ & -0.13 & [$-0.24, -0.02$] & -2.31 & .021 & -0.02 & [$-0.05, \phantom{-}0.00$] & -1.90 & .057 \\
\bottomrule
\end{tabular}
\end{table*}

\begin{table}[t]
\centering
\fontsize{9.5pt}{9.5pt}\selectfont
\caption{Spearman rank correlations between low-level stimulus features and the stimulus-level similarity measures ($n = 18$ stimuli; AUC and Adj.\ AUC averaged across orientations).}
\label{tab:feature_corr}
\renewcommand{\arraystretch}{1.2} 
\setlength{\tabcolsep}{4pt}
\begin{tabular}{
    l
    S[table-format=-.3]
    S[table-format=.3]
    S[table-format=-.3]
    S[table-format=.3]
}
\toprule
& \multicolumn{2}{c}{AUC} & \multicolumn{2}{c}{Adj. AUC} \\
\cmidrule(lr){2-3} \cmidrule(lr){4-5}
Feature & {$\rho$} & {$p$} & {$\rho$} & {$p$} \\
\midrule
Foreground pixel ratio &  .313 & .206 &  .191 & .448 \\
Content area           & -.476 & .046 & -.191 & .448 \\
Line density           &  .604 & .008 &  .205 & .414 \\
Perimeter              &  .430 & .075 &  .406 & .095 \\
Number of components   &  .471 & .049 &  .670 & .002 \\
Symmetry               &  .073 & .773 & -.077 & .760 \\
\bottomrule
\end{tabular}
\end{table}

\section{Discussion}
\label{sec:dsc}
\subsection{Visualization image}
\label{subsec:vsimg}

As shown in the FDMs of \figref{fig:visualization}, fixation locations in the animal stimulus images are predominantly concentrated around the head. This finding aligns with numerous studies indicating a tendency for gaze to focus on the face \citep{rousselet2003is,palermo2007are,crouzet2010fast}. Additionally, a portion of the fixations transitions from the head to the body. These fixation patterns remain consistent across all animal stimuli, including the rotated versions at $90^{\circ}$ and $180^{\circ}$. However, in the Elephant stimulus, fixations are concentrated not only on the head but also notably on the legs, a trend that becomes more pronounced at the $180^{\circ}$ rotation. This trend might be attributed to the exaggerated depiction of the Elephant's toenails, which were drawn larger. As a result, these oversized toenails might have been perceived as significant body parts, akin to facial features, leading to their misidentification as the head in the Elephant $180^{\circ}$, especially due to their placement in the upper part of the image.

In the non-animal stimuli, there is no common structure, such as the head, body, and legs, that is found in animal stimuli. As a result, the areas of concentrated fixation likely varied for each non-animal stimulus image. However, fixations were predominantly concentrated near the image's center for all non-animal stimuli, including their rotated versions, with the gaze generally spreading out from this central point. This pattern emerges because non-animal stimuli lack specific focal points, like faces in animal stimuli, that naturally guide the gaze. This suggests that this shift of gaze towards the periphery represents a search for functional areas following the initial center bias \citep{tseng2009quantifying,ohad2020what}. However, for the Truck and Revolver stimuli, fixations were more distributed across the entire figure, indicating that these images' figures occupied smaller areas, allowing the entire figure to be captured with minimal eye movement.

In the meaningless shape stimuli, there is a lack of both a common and functional structure. Consequently, the areas of concentrated fixation were likely to vary across each stimulus image. However, for all meaningless shape stimuli, fixations predominantly clustered near the center of the image. This can be reasonably interpreted as a significant manifestation of the center bias. Additionally, this indicates that eye movements were more scattered, reflecting uncertainty about where to direct the gaze.

On the other hand, the results of the PPF visualizations showed that potential values were typically higher in regions with more complex lines, regardless of the stimulus image. Additionally, high potential values were observed in parts of the contours, which are crucial for discerning the shapes' appearance.

\subsection{Similarity and variability score}
\label{subsec:sv_score}

As depicted in \figref{fig:similarity_category}(a), the similarity scores for animal stimuli display a sigmoidal pattern characterized by a sharp rise within approximately 1 second. \figref{fig:similarity_stimulus} further confirms this pattern for individual stimuli. However, the slope of the Elephant's similarity scores is notably less steep; accordingly, its Adj. AUC in \tabref{tab:results_compact} is considerably lower than those of the other animal stimuli, indicating that the exaggerated depiction of the toenails redirected the gaze towards the legs. Consequently, the similarity score appears to be susceptible to variations in sketch representation.

The similarity scores for non-animal stimuli, as presented in \figref{fig:similarity_category}(b), exhibit a sigmoidal pattern, albeit with a less steep slope compared to those of the animal stimuli. The middle row of \figref{fig:similarity_stimulus} illustrates that the similarity scores for the Truck and Revolver stimuli generally follow a sigmoidal pattern with higher values, whereas other non-animal stimuli show linear trends with lower values. This distinction is also evident in \tabref{tab:results_compact}, where the AUC values for Truck and Revolver are markedly higher than those for the remaining non-animal stimuli. In \figref{fig:visualization}, besides having a smaller content area, the Truck and Revolver stimuli display a relatively higher level of detail in patterns and textures compared to other non-animal stimuli. Therefore, similar to the animal stimuli, the similarity scores are substantially influenced by variations in sketch representation.

As illustrated in \figref{fig:similarity_category}, the similarity scores for the meaningless shape stimuli remain consistently low and follow a linear pattern. The bottom row of \figref{fig:similarity_stimulus} shows no variation in this linear trend, although the scores vary slightly between stimuli. This is further confirmed by \tabref{tab:results_compact}, where the Adj. AUC values for meaningless shape stimuli are near zero, in clear contrast to those of the animal stimuli.

Note that all categories show a non-trivial similarity score of approximately $0.52-0.55$ at viewing onset ($t=0$), consistent with the baseline mechanism described in the \nameref{subsec:similarity_score_metric} section. The Adj. AUC factors out this stimulus-dependent baseline and isolates the subsequent change in similarity attributable to the viewing process.

These qualitative observations were confirmed by the LMM analysis. As shown in \tabref{tab:pairwise}, pairwise comparisons revealed that the AUC of the meaningless shape category was significantly lower than those of both the animal and non-animal categories (both $p<.001$, Bonferroni-corrected), with large effect sizes ($d=4.41$ and $d=5.28$, respectively). However, the difference between animal and non-animal did not reach significance for the AUC after Bonferroni correction ($p=.108$). This non-significance is likely attributable to the influence of sketch detail, as the Truck and Revolver stimuli inflated the non-animal category mean. Indeed, when these two stimuli were excluded, the animal vs. non-animal comparison became significant ($p<.001$, $d=2.42$). In contrast, the Adj. AUC yielded significant differences for all three pairwise comparisons (all $p<.001$, Bonferroni-corrected), with large effect sizes (all $|d|>1.0$), indicating that baseline adjustment effectively reduces the influence of low-level sketch properties on the similarity scores.

\tabref{tab:feature_corr} summarizes the correlations between the stimulus features and the similarity measures. The raw AUC correlated substantially with line density ($\rho=.60$, $p=.008$) and content area ($\rho=-.48$, $p=.046$), quantitatively confirming that stimuli with small, densely drawn figures such as Truck and Revolver inflate the raw AUC. For the Adj. AUC, these correlations were markedly attenuated ($\rho=.21$ and $-.19$, respectively), and the symmetry score, although differing across categories, showed no association with either measure. These results support the claim that baseline adjustment reduces the influence of low-level stimulus properties. Two residual associations warrant attention. First, the number of connected components correlated with the Adj. AUC across all 18 stimuli ($\rho=.67$); however, this feature is intrinsically tied to the category distinction itself, as all meaningless shape stimuli consist of a single connected contour, and the association disappeared when the analysis was restricted to the twelve meaningful stimuli ($\rho=.11$). Second, within the meaningful stimuli, the Adj. AUC retained moderate correlations with the foreground pixel ratio and perimeter, indicating that the amount of drawn contour may still contribute to the difference between the animal and non-animal categories. However, when the foreground pixel ratio and line density were added as covariates to the LMM, all three pairwise category differences in the Adj. AUC remained significant (all $p<.001$, Bonferroni-corrected) with coefficients of the same sign, and the animal versus non-animal difference was, if anything, larger ($-0.16$ vs. $-0.11$). Although line density itself was a significant predictor, indicating that low-level properties do contribute to the similarity scores, the category effect was not explained by the measured features.

Regarding the effect of rotated presentation, the LMM revealed some significant main effects of orientation (\tabref{tab:orientation}); however, no significant Category $\times$ Orientation interactions were observed (all $p>.05$). Furthermore, the coefficients for 90° and 180° were nearly identical for the AUC, showing no graded pattern proportional to the rotation angle. These results indicate that rotated presentation did not produce a category-specific cognitive load effect that could be captured by the similarity scores.

On the other hand, $V'$ presented in \figref{fig:total_variability} reveals a clear distinction across categories: the meaningless shape stimuli exhibit predominantly low $V'$ values, whereas the animal and non-animal stimuli show a wider spread. While the FDMs for non-animal and meaningless shape stimuli in \figref{fig:visualization}, excluding Truck and Revolver, appear visually similar, $V'$ discerns the distinctions. This contrast for the meaningless shape stimuli can be interpreted as indicating that participants could not consistently identify a focal point in these shapes, representing a strong manifestation of the center bias. A notable exception was the elevated $V'$ for MS3 $90^{\circ}$. One possible explanation that warrants dedicated testing is that some participants may have associated these lines with familiar patterns or objects, a phenomenon known as pareidolia \citep{akdeniz2018neural,pepin2022processing}.

Additionally, the relationship between the AUC and $V'$ shown in \figref{fig:sv_relation}(a) illustrates the challenge in distinguishing between animal and non-animal stimuli based on these scores alone. As discussed above, this difficulty arises because both categories contain stimuli with varying levels of sketch detail. In contrast, the Adj. AUC in \figref{fig:sv_relation}(b) reduces the overlap between categories, and the LMM confirmed significant differences among all three categories (\tabref{tab:pairwise}), although some stimuli with atypical sketch detail, such as Truck, Revolver, and Elephant, visually remain in regions that overlap with other categories. In both panels, the meaningless shape stimuli form a distinct cluster in the lower-left region. This clustering indicates that, for meaningless shapes, fixations did not converge on regions that the PPF assigns high potential to, and that this held consistently across participants. The present design does not allow the cause to be identified: the meaningful and meaningless stimuli differ in low-level image features, contour complexity, and familiarity as well as in semantic content, and any of these could produce the observed separation. Dissociating these factors would require stimuli controlled for low-level visual properties or an explicit categorization task.

In summary, the overall trend observed in the similarity scores aligns with previous research \citep{awano2020psyochological,awano2022object}, particularly the rapid increase within 1 second for animal stimuli, which mirrors the speed of animal cognition \citep{kirchner2006ultra,zhu2013animal,zhu2016differential}. The LMM analysis confirmed that the three categories are statistically distinguishable, particularly when the baseline-adjusted measure is used to control for the influence of sketch representation. The similarity and variability scores did not capture the cognitive load associated with rotated presentation. The consistent clustering of meaningless shape stimuli in the low similarity and low variability region is the most robust pattern in the data, and it was obtained under free-viewing conditions without predefined category labels \citep{sarvadevabhatla2017object}.

\section{Conclusion and future work}
\label{sec:cncl}

This study reported the behavior of the similarity between FDM and PPF for simple sketches of three object categories under free viewing. During the eye-tracking experiment for generating FDMs, participants freely viewed stimulus images presented at their original and rotated orientations. The similarity between FDM and PPF was quantified as the similarity score, and a variability score was established to assess individual differences. Linear mixed models were used to statistically evaluate category and orientation effects.

Regarding the ability to distinguish object categories, the LMM analysis confirmed significant differences among the three categories. In particular, the baseline-adjusted measure (Adj. AUC) yielded significant pairwise differences among all three categories with large effect sizes, whereas the raw AUC did not distinguish between animal and non-animal categories due to the influence of sketch representation. A sensitivity analysis excluding stimuli with atypically high sketch detail further confirmed this interpretation. Regarding the effect of rotated presentation, the LMM revealed some significant main effects of orientation, but no graded pattern proportional to the rotation angle was observed. Moreover, no significant Category $\times$ Orientation interactions were found, indicating that the effect of rotation did not differ across categories. Regarding individual variation, the variability scores were generally low across all categories, with the meaningless shape stimuli showing particularly low values in both similarity and variability. The cause of this consistent trend for meaningless shapes cannot be determined from the present data, since low-level image features, familiarity, and semantic content are confounded across the stimulus categories. Further research focusing on meaningless shapes is also required, particularly by investigating factors such as the elevated variability observed for MS3 $90^{\circ}$, which may be related to pareidolia.

Finally, the participants in this study were limited to Japanese university students with relatively homogeneous knowledge and cultural backgrounds. Although the sensitivity power analysis confirmed adequate power for detecting the primary category differences, the modest sample size limits the ability to detect more subtle effects, such as within-category patterns or interactions. In addition, the present experiment employed a free-viewing condition without an explicit categorization task, and it remains an open question whether the observed patterns reflect categorization processes per se or spontaneous viewing behavior. Future studies involving larger and more diverse samples, as well as a comparison between free-viewing and explicit categorization conditions, are needed to enable finer-grained analyses and to assess the generalizability of the observed trends. In addition, although the amount of drawn contour was associated with the similarity measures, the category differences in the Adj. AUC remained significant after controlling for the measured low-level features. Nevertheless, stimulus sets that explicitly control these properties would allow a more complete dissociation of categorical and low-level contributions.

\section*{Acknowledgments}
This work was supported by JSPS KAKENHI Grant Number JP21K12098.

\section*{Declarations}
\paragraph{Ethical approval.} This study was performed in line with the principles of the Declaration of Helsinki and was approved by the Research Ethics Committee of Osaka University of Economics (approval number: 2020-B02). Written informed consent was obtained from all individual participants included in the study.
\paragraph{Conflicting interests.} The authors declared no potential conflicts of interest with respect to the research, authorship, and/or publication of this article.
\paragraph{Use of generative AI.} 
During the preparation of this work, the authors used ChatGPT (OpenAI) and Claude (Anthropic) to improve the readability and language of the manuscript. Draft sections originally written in the authors' native language (Japanese) were translated and refined into English with the assistance of these tools. In addition, these tools were used to support the development of the data analysis programs. After using these tools, the authors reviewed and edited the content as needed and take full responsibility for the content of this article.


\bibliographystyle{unsrtnat} 
\bibliography{references}

\end{document}